\documentclass[superscriptaddress,twocolumn]{revtex4}
\usepackage{amssymb}
\usepackage{amsfonts}
\usepackage{amsmath}
\usepackage{graphicx}
\usepackage{dcolumn}
\usepackage{bm}

\begin{document}

\title{Non trivial frames for $f(T)$ theories of gravity and beyond}
\author{Rafael Ferraro}
\email{ferraro@iafe.uba.ar}
\thanks{Member of Carrera del Investigador Cient\'{\i}fico (CONICET,
Argentina)}
\affiliation{Instituto de Astronom\'\i a y F\'\i sica del Espacio, Casilla de Correo 67,
Sucursal 28, 1428 Buenos Aires, Argentina}
\affiliation{Departamento de F\'\i sica, Facultad de Ciencias Exactas y Naturales,
Universidad de Buenos Aires, Ciudad Universitaria, Pabell\'on I, 1428 Buenos
Aires, Argentina}
\author{Franco Fiorini}
\email{franco@iafe.uba.ar}
\affiliation{Instituto de Astronom\'\i a y F\'\i sica del Espacio, Casilla de Correo 67,
Sucursal 28, 1428 Buenos Aires, Argentina}
\pacs{04.50.+h, 98.80.Jk}
\keywords{Teleparallelism, Born-Infeld, Cosmology}

\begin{abstract}
Some conceptual issues concerning $f(T)$ theories --a family of
modified gravity theories based on absolute parallelism-- are
analyzed. Due to the lack of local Lorentz invariance, the
autoparallel frames satisfying the field equations are evasive to
an \emph{a priori} physical understanding. We exemplify this point
by working out the vierbein (tetrad) fields for closed and open
Friedmann-Robertson-Walker cosmologies.
\end{abstract}

\maketitle

\section{Introduction: Teleparallelism in Weitzenb\"{o}ck spacetime}

There exists a general consensus that the description of the
gravitational field provided by general relativity (GR) is doomed
at scales of the order of the Planck length, where the spacetime
structure itself must be represented in terms of a quantum regime.
In the opposite extreme of the physical phenomena, GR also faces
an intriguing dilemma in connection with the late cosmic speed up
stage of the Universe. For these reasons, and for other purely
conceptual ones, GR has been the object of many extensions that
have tried to provide a more satisfactory description of the
gravitational field in the above mentioned extreme regimes. One of
the newest extended theories of gravity is the so-called $f(T)$
gravity, which is a theory formulated in a spacetime possessing
absolute parallelism \cite{Hehl1,Hehl2}. We started this idea in
Refs. \cite{Nos}-\cite{Nos3} by working out an ultraviolet
deformation of Einstein gravity. There we proposed a
Born-Infeld-like action with the aim of smoothing singularities,
namely the initial singularity of Friedmann-Robertson-Walker (FRW)
cosmological models. The proposal was successful in replacing the
initial singularity with an inflationary stage, so providing a
geometrical mechanism for the exponential increasing of the scale
factor without resorting to an inflaton. After that, the attention
was focused in low energy deformations of Einstein gravity, to
tackle those aspects of the cosmological evolution connected with
the late speed up stage of the Universe
\cite{fdet1}-\cite{fdet12}. Quite more recently, some fundamental
aspects of $f(T)$ theories, like the presence of extra degrees of
freedom and the lack of Lorentz invariance, had been addressed in
Refs.~\cite{miau}-\cite{fdetB2}.

In order to gain a deep insight into these and other features of
the $f(T)$ approach to modified gravity, it is mandatory to
enlarge the narrow number of physically relevant spacetimes
hitherto considered. In Section II we explain the lack of
invariance of $f(T)$ theories under local Lorentz transformation
of the field of vierbeins in a cosmological context, and discuss
about the meaning of this feature. In Section III, we work out the
proper vierbein for closed and open FRW cosmologies. The
discussion here stressed, in spite of its conceptual character,
leads to practical conclusions that will allow the comparison of
the cosmological consequences coming from different ways of
modified gravity. Finally, in Section IV we display the
conclusions.

The cornerstone of (four-dimensional) $f(T)$ theories is that
gravity can be described by providing the spacetime with a torsion
$T^{a}=de^{a}$, $a=0,...,3$, where $\{e^{a}\}$ is a vierbein (a
basis of the cotangent space) in a $4$-dimensional manifold
\footnote{$f(T)$-like gravities are non-trivial for dimensions
higher than two. The torsion tensor coming from an arbitrary
\emph{diad} $e^a(t,x)$ in $1+1$ dimensions has only two
independent components, $T^{t}_{\,\,\,t\,x}$ and
$T^{x}_{\,\,\,t\,x}$. However, the tensor $S_{\rho\mu\nu}$ that
takes part in the teleparallel Lagrangian of Eq. (\ref{Weitinvar})
is identically null. This property, in the light of Eq.
(\ref{divergence}), is consistent with the very known fact that
the Einstein-Hilbert action in two spacetime dimensions is just
the Euler characteristic class of the manifold.}. The vierbein
$\{e^{a}\}$ is the coframe of an associated basis $\{e_{a}\}$ in
the tangent space. If $e_{\mu }^{a}$ and $e_{a}^{\mu }$ are
respectively the components of the 1-forms $e^{a}$ and the vectors
$e_{a}$\ in a given coordinate basis, then the relation between
frame and coframe is expressed as
\begin{equation}
e_{\mu }^{a}\ e_{b}^{\mu }=\delta _{b}^{a}\ .  \label{coframe}
\end{equation}
Contracting with $e_{a}^{\nu }$ one also gets
\begin{equation}
e_{a}^{\nu }\ e_{\mu }^{a}=\delta _{\mu }^{\nu }\ .  \label{coframe2}
\end{equation}
The components $T_{\ \ \mu \nu }^{\lambda }$\ of the torsion tensor in the
coordinate basis is related to the 2-forms $T^{a}$ through the equation
\begin{equation}
T_{\ \ \mu \nu }^{\lambda }\equiv e_{a}^{\lambda }\ T_{\ \ \mu \nu
}^{a}=e_{a}^{\lambda }\,(\partial _{\nu }e_{\mu }^{a}-\partial _{\mu }e_{\nu
}^{a})\ .  \label{torsion}
\end{equation}
This means that the spacetime is endowed with a connection
\begin{equation}
{\Gamma }_{\mu \nu }^{\lambda }=e_{a}^{\lambda }\,\partial _{\nu
}e_{\mu }^{a}+\text{terms symmetric in}\,\, \mu \nu,  \label{Wei}
\end{equation}
(since $T_{\ \ \mu \nu }^{\lambda }\equiv {\Gamma _{\nu \mu
}^{\lambda }}-{\Gamma _{\mu \nu }^{\lambda }}$). The first term in
Eq.~(\ref{Wei}) is the Weitzenb\"{o}ck connection. The metric is
introduced as a subsidiary field given by
\begin{equation}
g_{\mu \nu }(x)=\eta _{ab}\ e_{\mu }^{a}(x)\ e_{\nu }^{b}(x)\ ,
\label{metric}
\end{equation}
where $\eta _{ab}=diag(1,-1,-1,-1)$. Eq.~(\ref{metric}) can be inverted with
the help of Eq.~(\ref{coframe}) to obtain
\begin{equation}
\eta _{ab}=g_{\mu \nu }(x)\ e_{a}^{\mu }(x)\ e_{b}^{\nu }(x)\ ,  \label{orto}
\end{equation}
which means that the vierbein is orthonormal.

The relation $T^{a}=de^{a}$ displays a remarkable analogy with the
electromagnetic field: the $e^{a}$'s play the role of potentials and
the $T^{a}$'s are the fields. The torsion $T^{a}$\ is invariant
under a gauge transformation $e^{a}\longrightarrow e^{a}+d\chi
^{a}$; the symmetric terms in the connection (\ref{Wei}) are the
imprint of such gauge transformation. Instead, the metric
(\ref{metric}) does not enjoy such gauge invariant meaning since it
is built from the ``potentials".

Teleparallelism uses the Weitzenb\"{o}ck spacetime, where the
connection is chosen as
\begin{equation}
{\Gamma }_{\mu \nu }^{\lambda }=e_{a}^{\lambda }\,\partial _{\nu
}e_{\mu }^{a}\ ,  \label{Wei1}
\end{equation}
Thus, the gauge invariance has been frozen. As a consequence of the
choice of the Weitzenb\"{o}ck connection (\ref{Wei1}), the Riemann
tensor is identically null. So the spacetime is flat: the
gravitational degrees of freedom are completely encoded in the
torsion $T^{a}=de^{a}$.

On other hand, the metric (\ref{metric}) does possess invariance
under local Lorentz transformations: $e^{a}\longrightarrow
e^{a^{\prime}} =\Lambda _{b}^{a^{\prime}}(x)\ e^{b}$; however the
torsion $T^{a}=de^{a}$ transforms as
\begin{equation}
T^{a}\longrightarrow T^{a^{\prime}} =\Lambda _{b}^{a^{\prime}} \,
T^{b}-e^{b}\wedge d\Lambda _{b}^{a^{\prime}} \ ,  \label{change}
\end{equation}%
which means that the exterior derivative of the vierbein is not
covariant under Lorentz transformations of the vierbein, unless
the Lorentz transformations be \textit{global}. This feature could
be healed by using a covariant exterior derivative to define
$T^{a}$ (i.e., by introducing a spin connection). This procedure
would restore the local Lorentz freedom of the vierbein. In 4
dimensions, this local freedom would reduce the 16 components
$e_{\mu }^{a}$ to only 10 physically relevant ones, but we should
add the new degrees of freedom encoded in the spin connection.
However, this strategy turned out to be inviable, see Ref.
\cite{fdetB1}.

In terms of parallelism, the choice of the Weitzenb\"{o}ck
connection has a simple meaning. In fact, the covariant derivative
of a vector yields
\begin{equation}
\nabla _{\nu }V^{\lambda }=\partial _{\nu }V^{\lambda }+{\Gamma
}_{\mu \nu }^{\lambda }V^{\mu } = e_{a}^{\lambda }\ \partial _{\nu
}(e_{\mu }^{a}V^{\mu })\equiv e_{a}^{\lambda }\ \partial _{\nu
}V^a\, .\label{transport}
\end{equation}
In particular, Eq.~(\ref{orto}) implies that $\nabla _{\nu
}e_{b}^{\lambda }=0$; so, the Weitzenb\"{o}ck connection is metric
compatible. In general, Eq.~(\ref{transport}) means that a given
vector is parallel transported along a curve if its projections on
the coframe remain constant. So, the vierbein parallelizes the
spacetime. Of course, this nice criterion of parallelism would be
destroyed if local Lorentz transformations of the coframe were
allowed in the theory.

Teleparallelism is a dynamical theory for the vierbein, which is
built from the torsion $T^{a}=de^{a}$. According to
Eq.~(\ref{metric}), a set of dynamical equations for the vierbein
also implies a dynamics for the metric. This dynamics coincides with
Einstein's dynamics for the metric when the teleparallel Lagrangian
density is chosen as \cite{Haya,Maluf}
\begin{equation}
\mathcal{L}_{\mathbf{T}}[e^{a}]=\frac{1}{16\pi G}\;e\;T\;,
\label{lagrangianT}
\end{equation}
where $e\equiv \det e_{\mu }^{a}=\sqrt{-\det (g_{\mu \nu })}$, and
\begin{equation}
T%
=S_{\ \mu \nu }^{\rho }T_{\rho }^{\ \mu \nu }. \label{Weitinvar}
\end{equation}
The tensor $S_{\ \mu \nu }^{\rho }$ appearing in the last equation
is defined according to
\begin{equation}
S_{\ \mu \nu }^{\rho }=\frac{1}{4}\,(T_{\ \mu \nu }^{\rho }-T_{\mu \nu }^{\
\ \ \rho }+T_{\nu \mu }^{\ \ \ \rho })+\frac{1}{2}\ \delta _{\mu }^{\rho }\
T_{\sigma \nu }^{\ \ \ \sigma }-\frac{1}{2}\ \delta _{\nu }^{\rho }\
T_{\sigma \mu }^{\ \ \,\sigma }\;.  \label{tensor}
\end{equation}
In fact, the Lagrangian (\ref{lagrangianT}) just differs from the
Einstein-Hilbert Lagrangian $\mathcal{L}_{\mathbf{GR}}=-(16\pi G)^{-1}\sqrt{%
-g}\ R$\ in a divergence
\begin{equation}
-e\ R[e^{a}]=e\ T-2\;\partial _{\nu }(e\;T_{\sigma }^{\ \ \sigma
\nu }\,)\;,  \label{divergence}
\end{equation}
where $R$ is the scalar curvature for the Levi-Civita connection.
When GR dresses this costume, the gravitational degrees of freedom
are gathered in the torsion instead of the Levi-Civita curvature.
It is a very curious and fortunate fact that both pictures enable
to construct a gravitational action with the same physical
content. However, it is remarkable that the Lagrangian
(\ref{lagrangianT}) involves just first derivatives of its
dynamical field, the vierbein. In some sense, the teleparallel
Lagrangian picks up the essential dynamical content of Einstein
theory without the annoying second order derivatives appearing in
the last term of Eq.~(\ref{divergence}). Such Lagrangian is a
better starting point for considering modified gravity theories,
since any deformation of its dependence on $T$ will always lead to
second order dynamical equations. On the contrary, the so-called
$f(R)$ theories lead to fourth order equations.

The teleparallel Lagrangian (\ref{lagrangianT}) can be rephrased in
geometrical language. Since $S_{\ \mu \nu }^{\rho }$ is antisymmetric in $%
\mu \nu $, then $e_{\rho }^{a}\ S_{\ \mu \nu }^{\rho }$ is a set of
four 2-forms $S^{a}$. Noticing that
\begin{equation*}
T_{\nu \mu}^{\ \ \rho }=g^{\rho \sigma }g_{\nu \lambda }\
e_{b}^{\lambda }\ T_{\ \mu \sigma }^{b}=\eta ^{df}e_{d}^{\rho
}\,e_{f}^{\sigma }\ \eta _{bc}\ e_{\nu }^{b}\ T_{\ \mu \sigma
}^{c},
\end{equation*}%
then we have%
\begin{equation*}
4\ S^{a}=T^{a}-\eta ^{ac}\ \eta _{bd}\ e_{c}\rfloor T^{b}\wedge
e^{d}+2\ e_{b}\rfloor T^{b}\wedge e^{a}\ ,
\end{equation*}%
where $e_{c}\rfloor T^{b}$ stands for the 1-form whose components are $%
e_{c}^{\sigma }T_{\sigma \mu }^{b}$. Thus, the teleparallel
Lagrangian density can be written as
\begin{equation}
\mathcal{L}_{\mathbf{T}}[e^{a}]=\frac{1}{16\pi G}\;\eta _{ab}\
S^{a}\wedge\, ^{\ast}T^{b}\;,  \label{lagrangianT2}
\end{equation}%
where $\ast $ is the Hodge star operator.

\section{The $f(T)^{\prime }s$ uncovered}

Analogously to the $f(R)$ scheme, a $f(T)$ theory replaces the
Weitzenb\"{o}ck invariant $T$ in Eq.~(\ref{lagrangianT}) with a
general function $f(T)$. So, the dynamics is described by the
action
\begin{equation}
\mathcal{I}=\frac{1}{16\pi G}\int d^{4}x\;\mathrm{e}\;f(T)+\int
d^{4}x\;\mathcal{L}_{\mathcal{M}}\;.  \label{action}
\end{equation}%
where $\mathcal{L}_{\mathcal{M}}$ is the matter Lagrangian density.
Undoubtedly, the whole family of actions gathered in (\ref{action})
constitutes a vast territory worth to be explored, specially when
one is aware that the dynamical equations arising by varying the
action (\ref{action}) with respect to the vierbein components
$e_{\mu }^{a}(x)$ are of second order. This distinctive feature
makes Weitzenb\"{o}ck spacetime a privileged geometric structure to
formulate modified theories of gravitation. In fact, the dynamical
equations for the vierbein are
\begin{eqnarray}
&&\mathrm{e}^{-1}\partial _{\mu }(\mathrm{e\ }S_{a}^{\ \ \mu \nu })\
\,f^{\prime }+e_{a}^{\ \ \lambda }S_{\rho }^{\ \ \nu \mu }T_{\ \ \mu \lambda
}^{\rho }\,\ f^{\prime }+  \notag \\
&&+S_{a}^{\ \ \mu \nu }\,\partial _{\mu }(T)\ \,f^{\prime \prime
}+ \frac{1}{4}e_{a}^{\ \ \nu }\,f=4\pi G\,\ \mathbb{T}_{a}^{\ \
\nu }, \label{dynamics}
\end{eqnarray}
where $\mathbb{T}_{a}^{\ \ \nu }=e_{a}^{\ \, \mu}\, T_{\mu }^{\nu
}$ refers to the matter energy-momentum tensor $T_{\mu \nu }$, and
the primes denote differentiation respect to $T$. These equations
tell how the matter distribution organizes the orientation of the
vierbein $e^{a}$ at each point, in such a way that the field lines
of $e^{a}(x)$ realize the parallelization of the manifold. After
this vierbein field is obtained, one uses the assumption of
orthonormality to get the metric (\ref{metric}). Instead, GR is a
theory for the metric; so it is invariant under local Lorentz
transformations of the vierbein. The equivalence between
Teleparallelism (\ref{lagrangianT}) and GR dynamics, expressed in
Eq.~(\ref{divergence}), implies that $T$ changes by a boundary
term under local Lorentz transformations. Because of this reason,
the teleparallel equivalent of GR does not provide the manifold
with a parallelization but only with a metric. On the contrary, in
a $f(T)$ theory the \textquotedblleft boundary term" in $T$ will
remain encapsulated inside the function $f$. This means that a
$f(T)$ theory is not invariant under local Lorentz transformations
of the vierbein. So, a $f(T)$ theory will determinate the vierbein
field almost completely (up to global Lorentz transformations). In
other words, a $f(T)$ theory will describe more degrees of freedom
than the teleparallel equivalent of GR. This is an important issue
in the search for solutions to the $f(T)$ dynamical equations,
since every two pair of vierbeins connected by a local Lorentz
transformation (i.e. leading to the same metric tensor) are
\emph{inequivalent} from the point of view of the theory. We will
address this topic in the next section by considering the
vierbeins that are suitable for closed and open FRW universes.

So far, the totality of the works alluding to $f(T)$ theories in
cosmological spacetimes deals with spatially flat FRW cosmologies.
This is not only because this geometry seems to be the more
appropriate for the description of the large scale structure of
the Universe, but also for technical reasons. In fact, one could
think that the starring field --the vierbein $\{e^{a}\}$, related
to the metric $g$ via $g=\eta _{ab}\ e^{a}\otimes e^{b}$ -- is the
naive square root of $g$. Thus, in a spatially flat FRW universe
described by the line element
\begin{equation}
ds^2=dt^2-a^2(t)\ (dx^2+dy^2+dz^2),
\end{equation}
one would replace in Eqs.~(\ref{dynamics}) the diagonal vierbein
\begin{equation}
e^{0}=dt,\,\,\,\,e^{1}=a(t)\,dx,\,\,\,e^{2}=a(t)\,dy,\,\,\,e^{3}=a(t)\,dz.
\label{tetplana}
\end{equation}%
This is really a good guess because the Eqs.~(\ref{dynamics}) become
a set of consistent dynamical equations for the scale factor $a(t)$.
Moreover, the Weitzenb\"{o}ck invariant for $T^{a}=de^{a}=\{0,\
\overset{.}{a}\, dt\wedge dx,\ \overset{.}{a}\, dt\wedge dy,\
\overset{.}{a}\, dt\wedge dz\}$ is
\begin{equation}
T=-6H^2,
\end{equation}
where $H=\dot{a}/a$ is the Hubble parameter. Additionally, it is
easy to check that the field equations (\ref{dynamics}) for the
vierbein (\ref{tetplana}) can be also obtained from a
(minisuperspace) reduced action constructed by replacing this
specific form of the Weitzenb\"{o}ck invariant in the general action
(\ref{action}).

It is not difficult to trace back the geometrical meaning of the
diagonal vierbein (\ref{tetplana}). Actually, the autoparallel
curves of flat Euclidean space are given by straight lines, which
can be generated by the coordinate basis $\partial _{i}$, whose dual
co-basis is just $dx_{i}$. Then, modulo a time-dependent conformal
factor, the frames describing the autoparallel lines are just
$dx_{i}$, as Eq.~(\ref{tetplana}) shows.

However, things\ are not so easy in the context of closed and open
FRW universes, whose line element can be described in
hyper-spherical coordinates as
\begin{equation}
ds^{2}=dt^{2}-k^{2}\,a^{2}(t)\,\ [d(k\psi )^{2}+\sin ^{2}(k\psi )\ (d\theta
^{2}+\sin ^{2}\theta \ d\phi ^{2})],  \label{metesferica}
\end{equation}
where $k=1$ for the closed universe and $k=i$ for the open
universe. Here, one is also tempted to think that the vierbein
that solves the dynamical equations (\ref{dynamics}) could have
the form
\begin{eqnarray}
&&e^{0^{\prime}}=dt,\notag\\
&&e^{1^{\prime}}=k\,a(t)\,d(k\psi ),\notag\\
&&e^{2^{\prime}}=k\,a(t)\,\sin (k\psi )\,d\theta,\notag \\
&&e^{3^{\prime}}=k\,a(t)\,\sin (k\psi )\,\sin \theta \,d\phi.
\label{naive}
\end{eqnarray}
In the teleparallel equivalent of GR, any choice of the vierbein
reproducing the metric is valid because of the local Lorentz
symmetry. On the contrary, the lack of this local invariance,
which is inherent to $f(T)$ theories, makes this naive choice to
be incompatible with the dynamical equations (\ref{dynamics}). In
other words, the vierbein (\ref{naive}) does not correctly
parallelize the spacetime. The symptom that the choice
(\ref{naive}) will not work is the form acquired by the
Weitzenb\"{o}ck invariant in such case, which turn out to be
\begin{equation}
T=2\,[(k\,a)^{-2}\,\cot^{2}(k\psi)-3H^2].
\end{equation}
This form of $T$ would be unable of giving a proper reduced
Lagrangian for the dynamics of the scale factor $a(t)$, as a
consequence of its dependence on $\psi $. This $\psi$-dependent
Weitzenb\"{o}ck invariant is not consistent with the isotropy and
homogeneity of the FRW cosmological models.

\section{Vierbeins for spatially curved FRW universes}

\subsection{Closed universes}

Let us discuss in detail the closed universe ($k=1$) with topology
$R\times S^{3}$. In order to parallelize the $S^{3}$ sphere, let
us consider $S^{3}$ as embedded in a 4-dimensional Euclidean space
with Cartesian coordinates $(X,\,Y,\,Z,\,W)$, so
\begin{equation}
X^{2}+Y^{2}+Z^{2}+W^{2}=1.  \label{sumcua}
\end{equation}%
At each point of the sphere there exists a Cartesian (canonical)
orthonormal coframe basis of the host Euclidean space, $\left\{
dX^{a}\right\} $, where $X^{a}$ stands for $(X,\,Y,\,Z,\,W)$. We
will rotate this coframe in such a way that one of the resulting
covectors be normal to the $S^{3}$ sphere, being the other three
covectors automatically tangent to $S^{3}$. This tangent vierbein
will prove to be the proper spatial part of the vierbein for the
closed FRW cosmology, in the sense that it will lead to consistent
dynamical equations for the scale factor of the closed universe. So,
let us introduce a smooth coframe field $\{\overcirc{E}^{a}\}$ on
the $S^{3}$ sphere by rotating the canonical frame $\left\{
dX^{b}\right\}$, i.e,
\begin{equation}
\overcirc{E}^{a}=R_{\;b}^{a}\ dX^{b}.  \label{newframe}
\end{equation}%
It is not difficult to verify that the matrix
\begin{equation}
R={\footnotesize \left(
\begin{array}{cccc}
Y & -X & -W & Z \\
W & -Z & Y & -X \\
-Z & -W & X & Y \\
X & Y & Z & W%
\end{array}%
\right) \;},
\end{equation}
constitute a local rotation on the sphere (it fulfills $\det R=1$,
and $R^{T}=R^{-1}$
on the $S^{3}$ sphere (\ref{sumcua})). Then, the rotated coframe (\ref%
{newframe}) turns out to be
\begin{eqnarray}
\overcirc{E}^{1} &=&Y\ dX-X\ dY-W\ dZ+Z\ dW  \notag \\
\overcirc{E}^{2} &=&W\ dX-Z\ dY+Y\ dZ-X\ dW  \notag \\
\overcirc{E}^{3} &=&-Z\ dX-W\ dY+X\ dZ+Y\ dW  \notag \\
\overcirc{E}^{4} &=&X\ dX+Y\ dY+Z\ dZ+W\ dW. \label{tetradaesfer}
\end{eqnarray}
Clearly, the covector $\overcirc{E}^{4}=(1/2)\
d(X^{2}+Y^{2}+Z^{2}+W^{2})$ is normal to the $S^{3}$ sphere; from
now on we shall focus on the tangent orthonormal coframe $\left\{
\overcirc{E}^{1},\ \overcirc{E}^{2},\ \overcirc{E}^{3}\right\} $.
For convenience, we will parametrize the $S^{3}$ sphere by using
hyper-spherical coordinates, which are related with the Cartesian
coordinates of the host Euclidean space in the usual manner
\begin{eqnarray}
X &=&\,\sin \psi \,\sin \theta \,\cos \phi   \notag \\
Y &=&\,\sin \psi \,\sin \theta \,\sin \phi   \notag \\
Z &=&\,\sin \psi \,\cos \theta \notag \\
W &=&\,\cos \psi .  \label{coordesfer}
\end{eqnarray}%
The angular coordinates range in the intervals $0\leq \phi \leq
2\pi $, $0\leq \theta \leq \pi $ and $0\leq \psi \leq \pi $. Thus,
we can expand the dreibein $\left\{ \overcirc{E}^{1},\
\overcirc{E}^{2},\ \overcirc{E}^{3}\right\} $ in the coordinate
basis $\left\{ d\psi ,\ d\theta ,\ d\phi \right\} $ to obtain
\begin{widetext}
\begin{eqnarray}
\overcirc{E}^{1} &=&-\cos \theta \ d\psi +\sin \psi \,\sin \theta
\ (\cos \psi \
d\theta -\sin \psi \,\sin \theta \ d\phi )  \notag \\
\overcirc{E}^{2} &=&\sin \theta \ \cos \phi \ d\psi -\sin \psi
\,[(\sin \psi \ \sin \phi -\cos \psi \ \cos \theta \ \cos \phi )\
d\theta +(\cos \psi \ \sin \phi +\sin \psi \ \cos \theta \ \cos
\phi )\ \sin
\theta \ d\phi ]\notag \\
\overcirc{E}^{3} &=&-\sin \theta \ \sin \phi \ d\psi -\sin \psi
\,[(\sin \psi \ \cos \phi +\cos \psi \ \cos \theta \ \sin \phi )\
d\theta +(\cos \psi \ \cos \phi -\sin \psi \ \cos \theta \ \sin
\phi )\ \sin \theta \ d\phi ].\notag \\ \label{autop1}
\end{eqnarray}
\end{widetext}
Finally, let us consider the vierbein
\begin{equation}
e^{0}=dt;\,\,\,e^{1}=a(t)\ \overcirc{E}^{1};\,\,\,e^{2}=a(t)\
\overcirc{E}^{2};\,\,\,e^{3}=a(t)\ \overcirc{E}^{3}.
\label{autop2}
\end{equation}
It is worth noticing that this vierbein can be directly obtained
from the naive vierbein (\ref{naive}) by means of a local rotation
whose Euler angles are $\psi ,\ \theta ,\ \phi $; in fact, both
frames are related via the Euler matrix
\begin{equation}
e^{a}=\mathcal{R}_{a^{\prime }}^{a}\ e^{a^{\prime }},
\label{rotacion}
\end{equation}
where
\begin{widetext}
\begin{equation*}
\mathcal{R}={\footnotesize \left(
\begin{array}{cccc}
1 & 0 & 0 & 0 \\
0 & 1 & 0 & 0 \\
0 & 0 & \cos \phi  & \sin \phi  \\
0 & 0 & -\sin \phi  & \cos \phi
\end{array}%
\right) \left(
\begin{array}{cccc}
1 & 0 & 0 & 0 \\
0 & \cos \theta  & \sin \theta  & 0 \\
0 & -\sin \theta  & \cos \theta  & 0 \\
0 & 0 & 0 & 1%
\end{array}%
\right) \left(
\begin{array}{cccc}
1 & 0 & 0 & 0 \\
0 & 1 & 0 & 0 \\
0 & 0 & \cos \psi  & \sin \psi  \\
0 & 0 & -\sin \psi  & \cos \psi
\end{array}
\right) }.
\end{equation*}
\end{widetext}
The fact that the naive vierbein (\ref{naive}) and (\ref{autop2})
are connected by a local rotation guarantees that the latter
actually describes the closed FRW metric given in
Eq.~(\ref{metesferica}) for $k=1$; i.e., it leads to the interval
\begin{equation*}
ds^{2}=dt^{2}-\,a^{2}(t)\,\ [d\psi^{2}+\sin ^{2}\psi\ (d\theta
^{2}+\sin ^{2}\theta \ d\phi ^{2})].
\end{equation*}
The applied procedure is successful because $S^{3}$, like all the
3-manifolds, is a parallelizable manifold, which means that it
accepts a globally well defined set of three smooth
($\mathcal{C}^{\infty }$) orthonormal vector fields, that serve as
a global basis of the tangent bundle $TM$.

The Weitzenb\"{o}ck invariant associated to the vierbein (\ref{autop2}) is%
\begin{equation}\label{invarcerrado}
T=6\,(a^{-2}-H^2),
\end{equation}%
which foretell that the vierbein (\ref{autop2}) will be adequate
to solve the dynamical equations. In fact, by replacing the
vierbein (\ref{autop2}) in the Eqs.~(\ref{dynamics}) one obtain
the modified version of Friedmann equation (for $a=0=\nu $):
\begin{equation}
12H^{2}f^{\prime }(T)+f(T)=16\pi G\rho . \label{friedcerrada}
\end{equation}
The equations for the spatial sector, ($a,\nu =1,2,3$), are equal
to
\begin{eqnarray}
&&4(a^{-2}+\dot{H})(12H^{2}f^{\prime \prime }(T)+f^{\prime }(%
T))-f(T)-\notag \\&&- 4 f^{\prime }(T)\ (2\dot{H}+3H^{2})=16\pi
Gp.  \label{espcerradas}
\end{eqnarray}
Note that Eq.~(\ref{friedcerrada}) is of first order in time
derivatives of the scale factor, irrespective of the function $f$.
Eqs. (\ref{friedcerrada}) and (\ref{espcerradas}) are two
differential equations for just one unknown function $a(t)$; so,
they are not independent. The way to see that this is indeed the
case, is to take the time derivative of (\ref {friedcerrada}) and
combine it with the conservation equation,
\begin{equation}
\dot{\rho}=-3H(\rho +p),  \label{densidad de energia}
\end{equation}
to obtain Eq.~(\ref{espcerradas}). Conversely, if the system
(\ref{friedcerrada}) and (\ref{espcerradas}) is consistent, then the
conservation of energy in the matter sector is given automatically
and Eq.~(\ref{densidad de energia}) holds.

\subsection{Open universes}

The Eq.~(\ref{rotacion}) also shows a way to find a proper vierbein
for the open universe: take the same rotation starting from the
naive vierbein (\ref{naive}) with $k=i$, but replace the Euler angle
$\psi $ with $i\psi $. The aspect of the so locally rotated frame is
now
\begin{widetext}
\begin{eqnarray}
\breve{E}^{1} &=&\cos \theta \ d\psi +\sinh \psi \,\sin \theta \
(-\cosh \psi \
d\theta +i\sinh \psi \,\sin \theta \ d\phi )  \notag \\
\breve{E}^{2} &=&-\sin \theta \ \cos \phi \ d\psi +\sinh \psi
\,[(i\sinh \psi \ \sin \phi -\cosh \psi \ \cos \theta \ \cos \phi
)\ d\theta +(\cosh \psi \ \sin \phi +i\sinh \psi \ \cos \theta \
\cos \phi )\ \sin
\theta \ d\phi ]  \notag \\
\breve{E}^{3} &=&\sin \theta \ \sin \phi \ d\psi +\sinh \psi
\,[(i\sinh \psi \ \cos \phi +\cosh \psi \ \cos \theta \ \sin \phi
)\ d\theta +(\cosh \psi \ \cos \phi -i\sinh \psi \ \cos \theta \
\sin \phi )\ \sin \theta \ d\phi ].\notag \\  \label{autop3}
\end{eqnarray}
\end{widetext}
Then the vierbein
\begin{equation}
e^{0}=dt;\,\,\,e^{1}=a(t)\ \breve{E}^{1};\,\,\,e^{2}=a(t)\
\breve{E}^{2};\,\,\,e^{3}=a(t)\ \breve{E}^{3}, \label{autop4}
\end{equation}
with the choice (\ref{autop3}), leads to the metric for the open FRW
cosmology:
\begin{equation}
ds^{2}=dt^{2}-a^{2}(t)\ [d\psi ^{2}+\sinh ^{2}\psi \ (d\theta
^{2}+\sin ^{2}\theta \ d\phi ^{2})].
\end{equation}%
In this case the Weitzenb\"{o}ck invariant is given by the
expression (compare with Eq. (\ref{invarcerrado}))
\begin{equation}\label{invarabierto}
T=-6\,(a^{-2}+H^2).
\end{equation}
The modified Friedmann equation arising from the vierbein
(\ref{autop4}) is again the Eq.~(\ref{friedcerrada}). Hence, the
equations coming from the spatial sector are (\ref{espcerradas}) but
with the change $a^{-2}\rightarrow -a^{-2}$ in the first term of the
expression.

\section{Final Remarks}

\begin{figure}[ht]
\includegraphics[scale=.50]{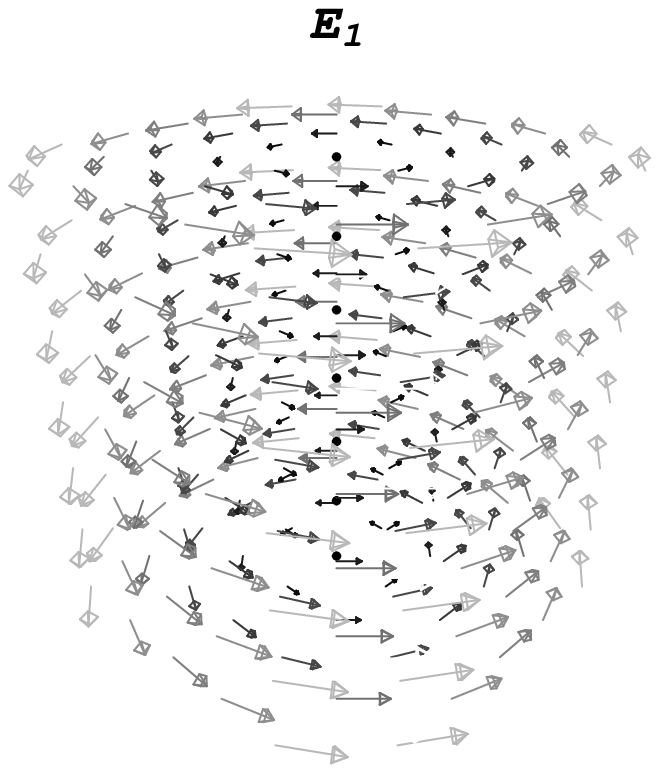}\newline
\includegraphics[scale=.50]{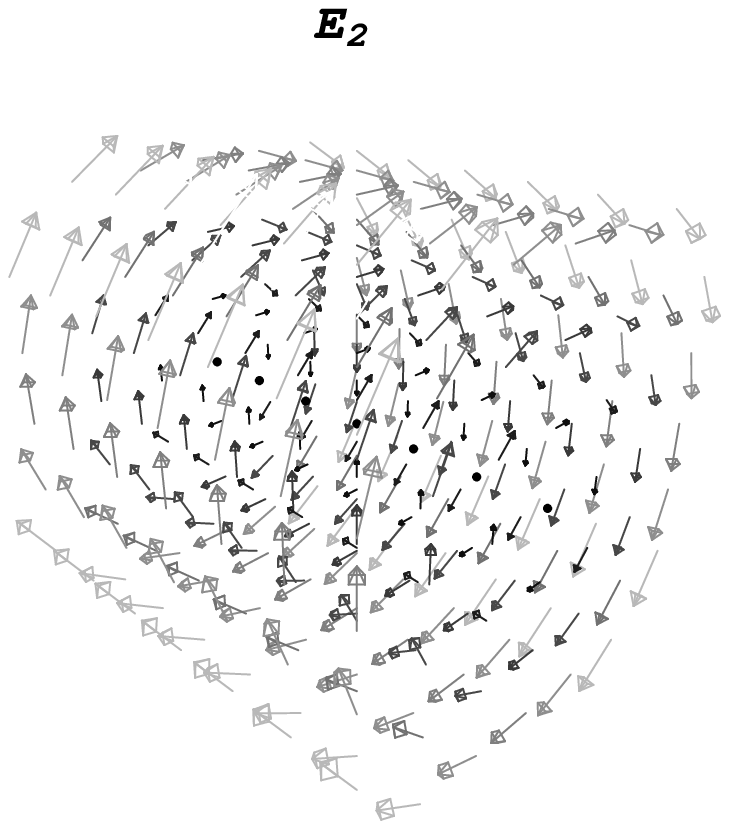}\newline
\includegraphics[scale=.50]{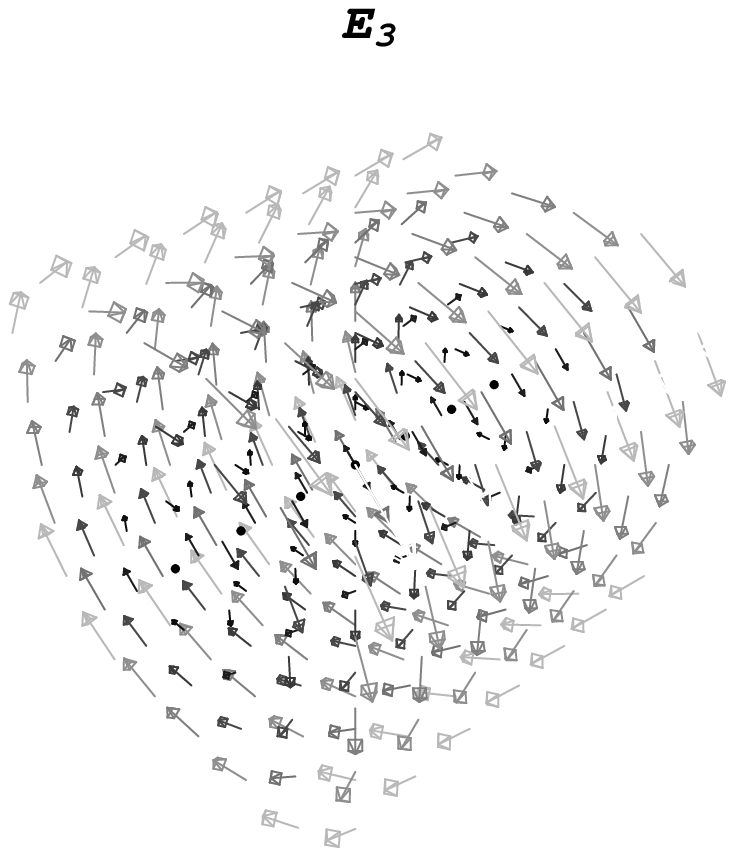}\newline
\caption{The parallel vector fields dual to the forms
(\protect\ref{autop1}), on the hyper-equator of the three
sphere.}\label{fig1}
\end{figure}

In the context of $f(T)$ theories, the spacetime structure is
materialized in the coframe field $\{e^a\}$ which defines an
orthonormal basis in the cotangent space $T_{p}^{\ast}M$ of the
manifold $M$ at each point $p\in M$. When $f(T)=T$, i.e., when one
consider general relativity in Weitzenb\"{o}ck spacetime, the
basis $\{e^a\}$ at two different points of the manifold are
completely uncorrelated, and it is not possible to define a global
smooth field of basis unambiguously. This is so because the theory
is invariant under the local Lorentz group acting on the coframes
$\{e^a\}$. In turn, when $f(T)\neq T$, local Lorentz rotations and
boosts are not symmetries of the theory anymore. Because of this
lack of local Lorentz symmetry, the theory picks up a preferential
global reference frame constituted by the coframe field $\{e^a\}$
that solves the field equations. In such case, the bases at two
different points become strongly correlated in order to realize
the parallelization of the manifold, as can be seen in the
\emph{fish shoal}-like pattern of Fig.~\ref{fig1}, showing the
vector fields $E_{1}$, $E_{2}$ and $E_{3}$, dual to the one-forms
of Eq.~(\ref{autop1}). In this figure, the coordinate $X_{4}$ was
set to zero, so the pictures represent the parallel vector fields
of $S^{3}$ in the hyper-equator defined by $\psi =\pi /2$,
immersed in three-dimensional Euclidean space.
The appearance of a preferred reference frame is a property coming
from the symmetries of the spacetime, and it is not ruled by the
specific form of the function $f(T)$. For instance, when one is
dealing with FRW cosmological spacetimes, the field of frames that
will lead to consistent field equations will be given by
(\ref{tetplana}), (\ref{autop2}) or (\ref{autop4}) depending on
whether the Universe is flat, closed or open respectively,
whatever the function $f(T)$ is. Additionally, these fields are
also valid in more general theories with absolute parallelism
which are not related with the $f(T)$ schemes. See, for instance,
Ref. \cite{Nos4}.

Let us finish with some additional remarks. Let be
$\{e^{\,a}(x)\}$ a vierbein field satisfying the vacuum Einstein
equations, i.e., $\{e^{\,a}(x)\}$ is a solution of
Eqs.~(\ref{dynamics}) with $f(T)=T$ and $\mathbb{T}_{\mu}^{\ \
\nu}=0$. Besides, suppose one can find a local Lorentz
transformation $\bar{e}^{a}(x)=\Lambda^a_b (x)\, e^{\,b}(x)$ such
that the Weitzenb\"{o}ck invariant $\bar{T}$ --which is invariant
under diffeomorphisms, but not under local Lorentz transformations
of the vierbein-- becomes globally zero. Then
$\{\bar{e}^{\,a}(x)\}$ is a solution not only for GR but for any
ultraviolet deformations of GR, i.e., for any theory described by
a function $f(T)$ verifying the condition
\begin{equation}  \label{condiciones}
f(T)=T+ \mathcal{O}(T^2),\,\,\,\text{i.e.},\,\,\, f(0)=0,\,\,\,
\text{and} \,\,\,f^{\prime}(0)=1.
\end{equation}
In fact, $\{\bar{e}^{\,a}(x)\}$ fulfills the Eqs.~(\ref{dynamics})
for any function $f(T)$ satisfying the high energy conditions
(\ref{condiciones}). This is because the prescription
(\ref{condiciones}) make the Eqs.~(\ref{dynamics}) to be exactly
the same that the GR ones whenever $T=0$. In other words, this
means that the original solution of the vacuum Einstein field
equations remains as a solution of the deformed $f(T)$ theories
described by the conditions (\ref{condiciones}). The so obtained
non-trivial coframe $\{\bar{e}^a(x)\}$ could serve as the starting
point to search for vacuum solutions of arbitrary infrared $f(T)$
deformations, like the ones considered in the literature in order
to explain the late time cosmic speed up of the Universe. This
practice will enable to further reduce the set of physically
relevant $f(T)$ models by using the well established post
Newtonian constraints \cite{chino}, \cite{turcos}, and hopefully,
will shed some light in the question of which are the extra
degrees of freedom hidden behind $f(T)$ gravities. We shall have
the opportunity to deal with these matters in a future work.

\acknowledgements We are particularly grateful to M. Bellini, E.
Calzetta and D. Mazzitelli for reading the original version of the
manuscript. This research was supported by CONICET and Universidad
de Buenos Aires.

\end{document}